\documentclass[aps,prd,nofootinbib,twocolumn,groupedaddress,showpacs]{revtex4}

\usepackage{graphicx}

\begin{document}

\title{Black-hole binary simulations: the mass ratio 10:1}

\author{Jos{\'e} A.~Gonz{\'a}lez$^{1,2}$, Ulrich Sperhake$^{1,3}\footnote{Electronic address: sperhake@tapir.caltech.edu}$, Bernd Br{\"u}gmann$^1$}
\affiliation{${}^1$ Theoretisch-Physikalisches Institut, Friedrich-Schiller-Universit\"at, 07743 Jena, Germany}
\affiliation{${}^2$ Instituto de F\a'{\i}sica y Matem\a'aticas, Universidad Michoacana de San Nicol\a'as de Hidalgo, Morelia, Michoac\a'an, M\a'exico}
\affiliation{${}^3$ Theoretical Astrophysics 350-17, California Institute of Technology, Pasadena, CA 91125}

\date{\today}

\begin{abstract}
We present the first numerical simulations of an initially non-spinning
black-hole binary with
mass ratio as large as $10:1$ in full general relativity. The binary
completes approximately 3 orbits prior to merger and radiates
$(0.415 \pm 0.017)~\%$ of the total energy and $(12.48\pm0.62)~\%$
of the initial angular momentum in the form of gravitational waves.
The single black hole resulting from the merger
acquires a kick of $(66.7\pm 3.3)~{\rm km/s}$ relative to the original
center of mass frame.
The resulting
gravitational waveforms are used to validate existing formulas for
the recoil, final spin and radiated energy over a wider
range of the symmetric mass ratio parameter $\eta=M_1M_2 / (M_1+M_2)^2$
than previously possible.
The contributions of $\ell > 2$ multipoles are found to visibly influence
the gravitational wave signal obtained at fixed inclination angles.
\end{abstract}

\pacs{04.25.D-, 04.25.dg, 04.30.Db}

\maketitle

\section{Introduction}

Following the breakthroughs of 2005
\cite{Pretorius2005a,Campanelli2006,Baker2006},
the numerical relativity community has constructed several independent
codes \cite{Pretorius2006, Baker2006a, Campanelli2006a,
Herrmann2006, Sperhake2006, Bruegmann2006a, Pollney2007,
Etienne2007a, Scheel2008}
and proceeded
at a breathtaking speed in generating new insights into the dynamics
of inspiralling and coalescing black-hole binaries (BBH) and the resulting
gravitational wave patterns \cite{Pretorius2007a}.
For example, numerical relativity has provided
vital information regarding the
kick or recoil in astrophysical mergers
\cite{Baker2006b, Gonzalez2007, Herrmann2007,
Koppitz2007, Campanelli2007, Gonzalez2007a,
Campanelli2007a, Bruegmann2007, Lousto2007, Healy2008},
simulations of spin flip and precession
phenomena \cite{Campanelli2006d, Campanelli2007} and the interpretation
of gravitational waveforms to be observed in the future
\cite{Baumgarte2006, Buonanno2006, Berti2007,
Ajith2007, Ajith2007a, Vaishnav2007}.
See also \cite{Shibata2006a, Anderson2007,
Etienne2007, Baiotti2008, Duez2008}
for binary simulations involving neutron stars.
This progress has come timely, as
earthbound gravitational wave detectors LIGO, VIRGO, GEO600 and TAMA
\cite{Abbott2007a, Beauville2007, Lueck2006, Tatsumi2007}
are now collecting data at or near the design sensitivity.
The combined
use of theoretical predictions and sophisticated data analysis
methods will be crucial in achieving the first direct detection
of gravitational waves and thus opening a new window to the universe.
%{\bf Refs?}.

Purpose of this work is to extend the range of mass
ratios probed by numerical relativity to $q\equiv M_1/M_2=10$ corresponding to
$\eta\equiv q/(1+q)^2=0.0826$. This mass ratio is of particular interest for
a variety of reasons.
First, studies of the supermassive black hole formation
history starting with light seed black holes predict a significant
fraction of mergers in the range $1\le q\le 10$ and, depending on
accretion details, the possibility of a peak near $q=5...10$
\cite{Sesana2007}.
Similarly,
detailed statistical
analysis of the mass distribution of supermassive galactic black holes
predicts that most mergers will occur in the range $3\le q\le 30$
\cite{Gergely2007}.
Finally, accurate numerical simulations
with $q=10$ will facilitate unprecedented comparisons with
approximative calculations based on post-Newtonian expansions
(cf.~\cite{Damour1998, Buonanno1999, Damour2001, Arun2004} as well as
\cite{Blanchet2006} for a review)
and extreme mass ratio predictions
based on perturbation theory and self force calculations
(see \cite{Mino1997, Poisson2004, Hughes2006, Damour2007c,
Detweiler2008, Hinderer2008, Yunes2008} and references
therein).
By considering, for example, an expansion in the
mass ratio parameter, a naive estimate of the error in
perturbative calculations would be of the order of $1/q^2$ or $1~\%$
relative to background quantities,
or $10~\%$ for quantities of first order in $1/q$.
While such comparisons are beyond the scope of this paper, we will
lay the foundation for future work by giving a detailed convergence
analysis of the numerical results including estimates for the uncertainties.

We will also use our results to probe recently published
formulas for calculating the final spin and recoil resulting from the
coalescence of two black holes with given initial physical parameters.
Spin measurements of black holes via direct astrophysical observations
have so far provided information about several individual holes
\cite{Wilms2001, Elvis2002, Davis2006, Middleton2006,
McClintock2006, Shafee2006, Wang2006} but appear as yet to be insufficient
for constructing reliable spin-distribution functions.
The community has therefore pursued the alternative path of using
theoretical predictions
in the context of the growth history and accretion processes of
the holes \cite{Hughes2003, Gammie2003, Shapiro2004, Volonteri2004,
King2006, Berti2008, King2008}. It is important for the modeling of
the individual binary mergers in such simulations to have available
mappings between the initial parameters of the binary and the final
spin and recoil of the post-merger remnant.
A better understanding of the distribution function of the black-hole
recoil also generates a great deal of interest in its own right.
In particular, there remain open questions as to how generically
recoil velocities of thousands of km/s
result from astrophysically realistic binary mergers.
Such large recoil would not only result in
intergalactic populations of black holes but also
affect the central structure of galaxies
and put severe constraints on possible scenarios of the black hole
formation history
\cite{Redmount1989, BoylanKolchin2004, Haiman2004, Madau2004, Merritt2004,
Libeskind2006, Blecha2008}. For recent discussions on
direct observational signatures of recoiling
black holes, see also \cite{Komossa2008, Komossa2008a, Menou2008}.

Existing formulas predicting the kick and final spin as functions of the
initial parameters of the binary are based on
simulations using mass ratios
$1\le q \le 4$ or $0.25 \ge \eta \ge 0.16$ and, in the case of
Baker {\em et al.}~\cite{Baker2008a}, using also $q=6$ or $\eta=0.122$
for the final spin. Our results clarify
the validity of these formulas in the case of initially non-spinning
holes. We further analyze the multipolar structure of the resulting
gravitational waveforms and illustrate the significance of
the higher order multipoles in the gravitational wave signal.

The paper is organized as follows. We describe
in Sec.~\ref{sec: numerics} the numerical simulations
together with a calibration of their uncertainties.
In Sec.~\ref{sec: GWs}, we analyze the gravitational waveforms
generated in the inspiral with regard to the total amount of
energy, linear and angular momentum radiated and test the validity
of existing fitting formulas for final spin and recoil. We further
demonstrate that a significant fraction of the energy is radiated
in higher order multipoles which implies that gravitational wave
signals differ significantly from the typical quadrupole
shape presented in most of the numerical relativity literature.
We conclude in Sec.~\ref{sec: conclusions} and discuss possible
future studies based on the data presented in this work.

\section{Numerical simulations}
\label{sec: numerics}

We have performed our simulations using the {\sc bam} code as
described in \cite{Bruegmann2006a,Bruegmann2004}. In order to obtain sufficient
accuracy for this demanding type of simulations,
we use sixth order discretization
for the spatial derivatives \cite{Husa2007a} and fourth order
accurate integration in time. Initial data are provided by the
spectral solver described in Ref.~\cite{Ansorg2004}. Gravitational
waves are calculated in the form of the Newman-Penrose scalar
$\Psi_4$ according to the procedure described in Sec.~III in
\cite{Bruegmann2006a} (but see \cite{Nerozzi2008} for an
alternative method to extract $\Psi_4$).

The model we are focusing on in this work represents a black-hole
binary with initial parameters as given in Table \ref{tab: model}.
\begin{table*}[t]
  \caption{Initial parameters and main results
           of the 10:1 mass ratio simulation studied in this work.
           $m_{1,2}$ and $M_{1,2}$ are the bare masses and black hole masses
           respectively. The Bowen-York parameters for the tangential linear
           momentum $P$ and the coordinate separation $D$ are normalized
           with respect to the total black hole mass $M=M_1+M_2$.
           Radiated energy and angular momentum are normalized with respect
           to their total values for the spacetime. Finally, we give
           the dimensionless spin and kick parameter of the merged hole.}
  \begin{ruledtabular}
  \begin{tabular}{ccccccccccc}
  $m_1$     & $m_2$      & $M_1$     & $M_2$     & $M_{\rm ADM}$ & $P/M$     & $D/M$     & $\frac{E_{\rm rad}}{M_{\rm ADM}}$   & $\frac{J_{\rm rad}}{J_{\rm tot}}$    & $j_{\rm fin}$         & $v_{\rm kick}$ \\
  \hline
  2.4831    & 0.2303     & 2.5       & 0.25      & 2.7381        & 0.0415    & 7.0       & $(0.415 \pm 0.017)\%$               & $(12.48 \pm 0.62)\%$                 & $0.259 \pm 0.003$     & $66.7 \pm 3.3~{\rm km/s}$\\
  \end{tabular}
  \label{tab: model}
  \end{ruledtabular}
\end{table*}
We follow the convention of \cite{Sperhake2007} and normalize initial
parameters relative to the total black hole mass $M=M_1+M_2$ and
dimensional
diagnostic quantities by their total initial values;
the Arnowitt-Deser-Misner (ADM) mass $M_{\rm ADM}$ and the
total initial angular momentum $J_{\rm ini}$. The initial
tangential linear momentum $P$ has been calculated from Eq.~(65)
of \cite{Bruegmann2006a} which gives a 3rd order post-Newtonian
estimate for the momentum of a quasi-circular configuration.
We calculate the number of orbits completed by this configuration
from the waveform as described in Sec.~III C of \cite{Sperhake2007}
and obtain about 3 orbits and 6 wave cycles.

The mass ratio $q = 10$ does not require any changes in the
construction of initial data. However, let us point out one important issue
that affects the puncture evolution method. We use the ``$000$'' gauge
advection choice, that is we evolve the shift according to $\partial_0
\beta^i = \frac{3}{4}B^i$ and $\partial_0 B^i = \partial_0
\tilde{\Gamma}^i-\eta_s B^i$ with $\partial_0 = \partial_t - \beta^i
\partial_i$. In second order form the shift condition is
\begin{equation}
  \partial_0^2 \beta^i = 
  \frac{3}{4} \partial_0 \tilde{\Gamma}^i - \eta_s\partial_0\beta^i,
\end{equation}
from which it is immediate that the physical dimension of the shift
damping parameter $\eta_s$ is
\begin{equation}
   [\eta_s] = 1/M.
\end{equation}
(We have added the label ``s'' to distinguish $\eta_s$ from the
mass ratio $\eta$ used elsewhere.)
The parameter $\eta_s$ was introduced to control the dynamics of the shift
vector \cite{Alcubierre2001,Alcubierre2003b}; in
particular it influences the degree of slice stretching that develops
near the black holes during dynamic gauge evolution, but $\eta_s$ also
affects the drift of coordinates near the outer boundary. 
Only certain
values of $\eta_s$ lead to stable evolutions. This also applies to
evolutions in the moving puncture framework, see e.g.\
\cite{Bruegmann2006a} for a discussion of some of the $\eta_s$
dependence found in our evolutions, and \cite{vanMeter2006} for related
discussions.

The issue that arises for $q=10$ is that $\eta_s$ is chosen to be a
global constant, say $\eta_s=2.0/M$, but the effect of $\eta_s$ on the
slice stretching near the black holes is different if the black hole
masses are different.  Assuming that $\eta_s=2.0/M$ is a good choice
near the black holes for equal mass, $M_1 = M_2 \approx M/2$, then
increasing $M_1$ by a factor of 5
means that $\eta_s$ has to be replaced by $\eta_s/5$ to obtain the
same amount of slice stretching. Similarly, if $M_2$ is reduced,
then $\eta_s$ should be correspondingly enlarged.

For these reasons the standard choice of $\eta_s=2.0/M$ made in
\cite{Bruegmann2006a} did not work for $q=10$ evolutions; the
effective $\eta_s$ near the black holes did not lead to stable and
accurate evolutions. A tell-tale sign for $\eta_s$ being too large is
if the orbits drift outwards rather than spiral inwards, which is
accompanied by a loss of convergence with time. Numerical experiments
revealed that $\eta_s$ can be chosen in a certain interval without
loss of convergence, which warrants a detailed study in the
future. For the present purpose it is sufficient to note that choosing
\begin{equation}
  \eta_s = 1.375/M
\end{equation}
works near both black holes even for $q=10$.
We plan to investigate possible benefits of a non-constant, position dependent
$\eta_s$ (as already suggested in \cite{Alcubierre2001,Alcubierre2003b})
in a future publication.

A computational issue arising for increasing mass ratios is that such
simulations typically require more computer time than simulations for
comparable masses. The resolution requirement and the time step size
is determined by the smaller mass, while the physical time required
for one orbit increases with the total mass. We illustrate this
by giving rough estimates for the number of computational time steps
required for simulating the last $3$ orbits prior to merger.
In the equal-mass case $q=1$, reasonable accuracy can be obtained
by using a resolution $h=M/48$ on the finest refinement level
corresponding to about $25,000$ time steps for the last $3$ orbits.
For the present simulation with $q=10$, on the other hand, we obtain
the same number of orbits after about $250,000$ time steps in the
medium resolution case labeled $N=68$ below. It is imperative,
therefore, to maintain high numerical accuracy over a longer
evolution time as we increase $q$.
A similar increase in computational demands was noted
by Lousto \& Zlochower \cite{Lousto2008}, who use an additional
refinement level for spinning binaries with $q=4$ (see their Sec.~III).

For these reasons, it is not permissible to infer information on the
numerical uncertainties of our simulations by merely extrapolating
convergence studies of equal- or mildly unequal-mass binaries
as obtained for example in \cite{Bruegmann2006a, Berti2007}.
Instead, we need to study convergence of the present scenario
using three resolutions. In the notation of
Sec.~VI A of \cite{Bruegmann2006a}, our grid setup is given by
$\chi_{\eta_s=1.375}[3\times N:6\times 2N:6]$, where the number of grid
points is $N=60$, $68$ and $76$ for the low, medium and high
resolution runs, respectively.
\begin{figure}[b]
  \includegraphics[angle=-90,width=250pt]{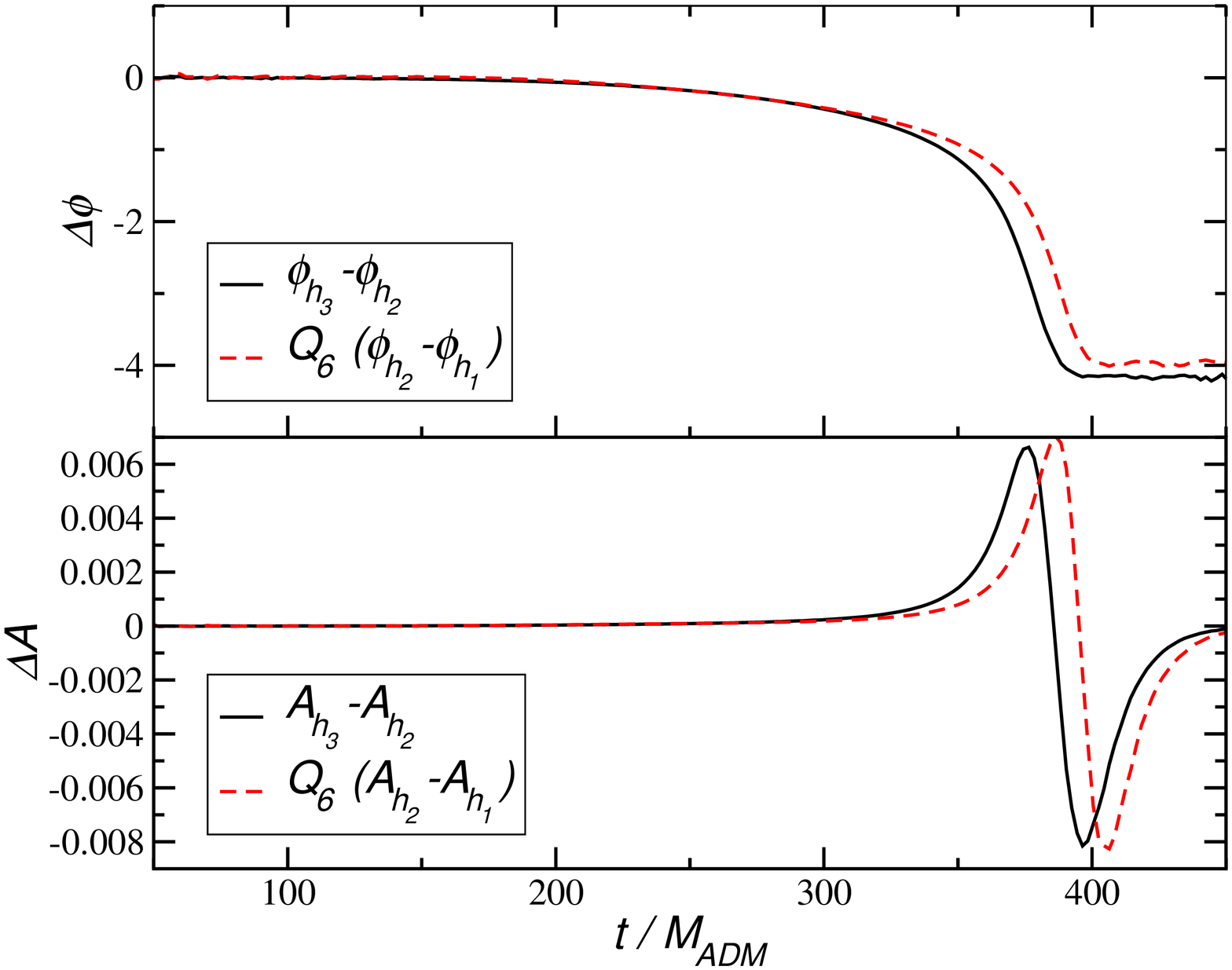}
  \includegraphics[angle=-90,width=250pt]{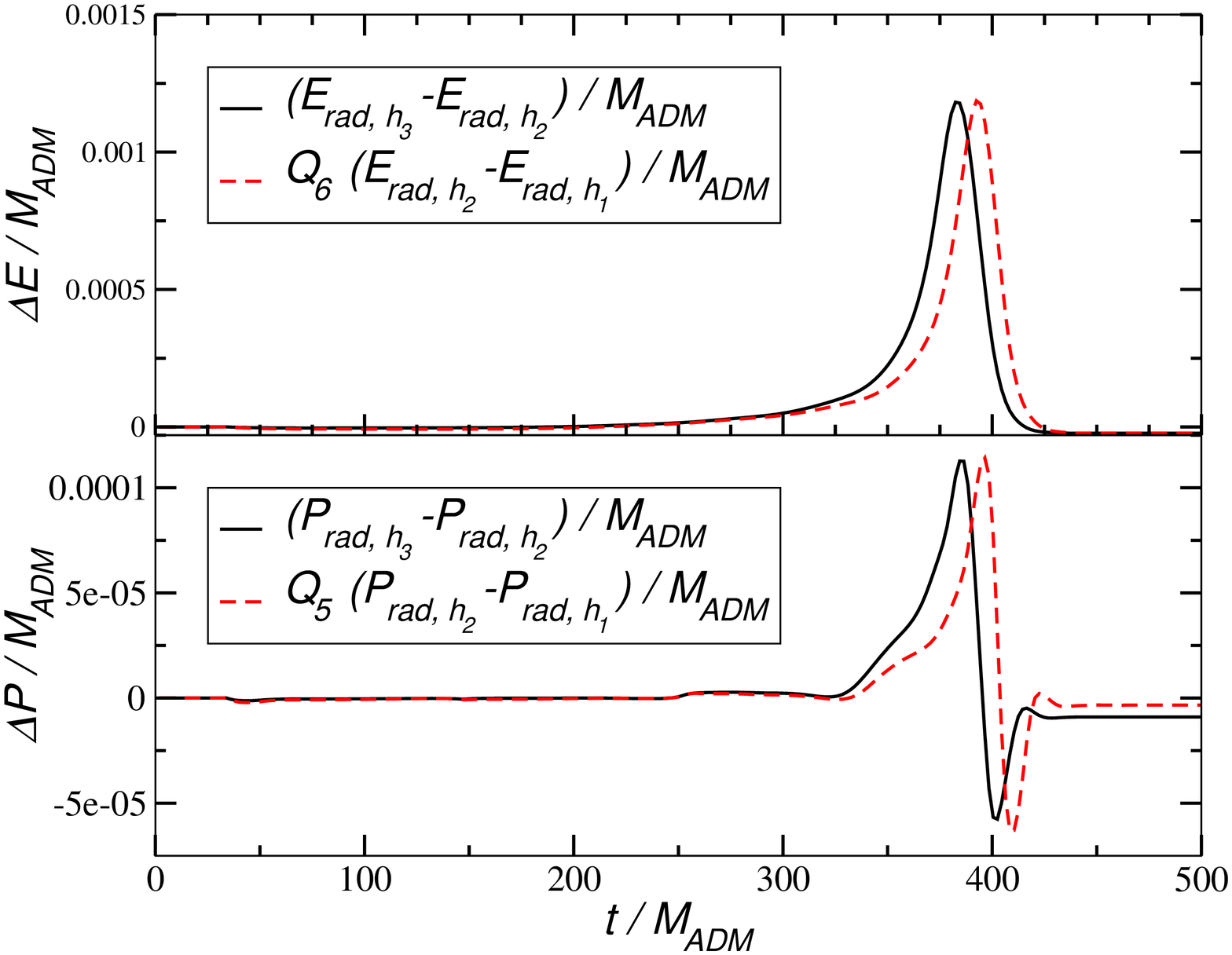}
  \caption{Convergence plots for the amplitude $A$ and phase $\phi$
           of the $\ell=2$, $m=2$ mode of the Newman Penrose
           scalar $\Psi_4$ (upper panels) as well as the radiated
           energy $E$ and linear momentum $P$ (lower panels). The
           scaling factors for fifth and sixth order convergence
           are $Q_5=2.04$ and $Q_6=2.30$.
          }
  \label{fig: convergence}
\end{figure}
In Fig.~\ref{fig: convergence} we show the resulting
convergence analysis obtained for
the $\ell=2$, $m=2$ mode of the Newman-Penrose scalar and the
radiated energy $E$ and linear momentum $P$ extracted at
$r_{\rm ex}=36.5~M_{\rm ADM}$. For this purpose, we have decomposed
the wave signal according to
\begin{equation}
  \psi_{22} = A e^{i\phi},
\end{equation}
and studied amplitude $A$ and phase $\phi$ separately. We observe
$6^{\rm th}$ order convergence for both quantities and the total
radiated energy. For the linear momentum $P = \sqrt{P_x^2+P_y^2}$,
the convergence is closer to $5^{\rm th}$ order. We estimate the
error due to discretization
by comparing the high resolution solution with
the Richardson extrapolated result assuming $4^{\rm th}$ order
convergence. We use this more conservative $4^{\rm th}$ order
extrapolation to account for $4^{\rm th}$ order accurate ingredients
in the {\sc bam} code.
The uncertainties thus obtained are about $5~\%$ for the
phase, $1~\%$ for the amplitude and radiated energy and $3~\%$
for the recoil and radiated angular momentum.
We emphasize that no alignment in phase or time
of the wave signal has been applied for this analysis.

We similarly determine the error arising from extracting the waves
at finite radius. The extraction radii available for this analysis are
$r_{\rm ex}=18.3$, $27.4$ and $36.5~M_{\rm ADM}$.
For the radiated
angular momentum, for example, we obtain at these
radii $J_{\rm rad}=12.06$, $12.31$
and $12.46~\%$ respectively of $J_{\rm ini}$. These values are
well modeled by a function $a_0 + a_1/r_{\rm ex}$
which gives us a relative
error of about $3.5~\%$ for the value extracted at the largest
radius $36.5~M_{\rm ADM}$. Similar results are obtained for
the other radiated quantities.
We investigate the phase error of the $22$ mode
due to the extraction radius in the
same way. We consider the phase as a function of
retarded time $u=t-r_{\rm ex}$ and fit for each value of $u$
\begin{figure}[b]
  \includegraphics[angle=-90,width=250pt]{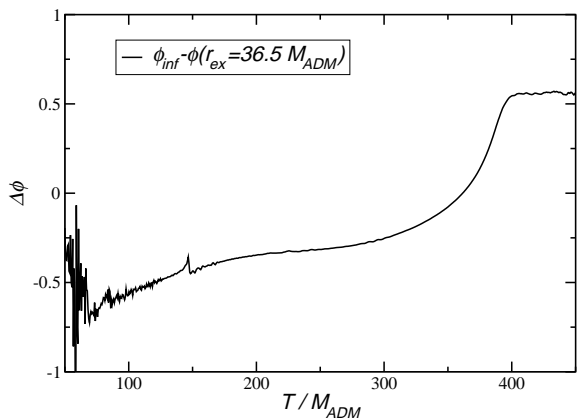}
  \caption{Difference in phase of the $\ell=2$, $m=2$ mode
           extrapolated to infinite extraction radius and
           extracted at $r_{\rm ex}=36.5~M_{\rm ADM}$.
          }
  \label{fig: phirex}
\end{figure}
the function $\phi(u, r_{\rm ex})= \phi_0(u) + \phi_1(u) / r_{\rm ex}$.
The difference between the extrapolated phase and the result obtained
at $r_{\rm ex}=36.5~M_{\rm ADM}$ is shown in Fig.~\ref{fig: phirex}.
The accumulated phase error after coalescence
is about $\Delta \phi = 1.5$ rad corresponding
to a relative error of about $2~\%$.
The magnitude of these errors obtained at comparatively small
extraction radii seems to suggest that for $q=10$ certain near zone
effects are reduced compared to $q=1$, although this is to be examined
more closely in future work.

We combine the uncertainty estimates due to discretization and
extraction radius assuming standard error propagation and adding the
squares of the individual errors. We thus obtain error estimates
of about $4~\%$ for the wave amplitude and
radiated energy, $5~\%$ for the radiated momenta
and $5.5~\%$ for the phase. Note that most of the phase
error builds up during the late stages of the inspiral and merger, so
that phase uncertainties for performing a comparison with post-Newtonian
results will be smaller than the estimate given here.
We summarize the results for the radiated energy, momenta and
final spin in Table \ref{tab: model} with error bars
based on the above analysis.

\section{Gravitational wave emission}
\label{sec: GWs}

In this section, we study in more detail the gravitational wave
emission of the binary. We also discuss the resulting values
for radiated energy, final spin and recoil in the context of
general formulas suggested in the literature.

\subsection{Comparison with phenomenological formulas}
\label{sec: fitting}

All theoretical modeling of the mass and spin evolution of
black holes in the context of their merger history
requires a mapping between initial
and final black hole parameters for each individual merger.
The generation of such mappings has recently become an industrious
area of research, largely because of the breakthroughs in numerical
relativity which make it possible now to simulate black-hole binaries
accurately through inspiral, merger and ringdown.
Results are currently available for a relatively small subset of the
parameter space only, and have resulted in various efforts to
``extrapolate'' to wider ranges of the input parameter space using
(semi-)analytic methods
\cite{Buonanno2007, Rezzolla2007, Boyle2007a,
Rezzolla2007a, Rezzolla2007b, Boyle2007b, Tichy2008}.
The fitting formulas
thus generated, have been relatively well tested in the
regime of binaries with nearly equal mass or in the test particle
limit, but in the regime in between, corresponding to
a symmetric mass ratio in the range $\eta=0.05...0.12$, accurate
data have as yet not been available. Here, we fill this gap and
test existing predictions for the case of a non-spinning binary
with $\eta=0.0826$ or $q=10$.
\begin{figure}[b]
  \includegraphics[angle=-90,width=250pt]{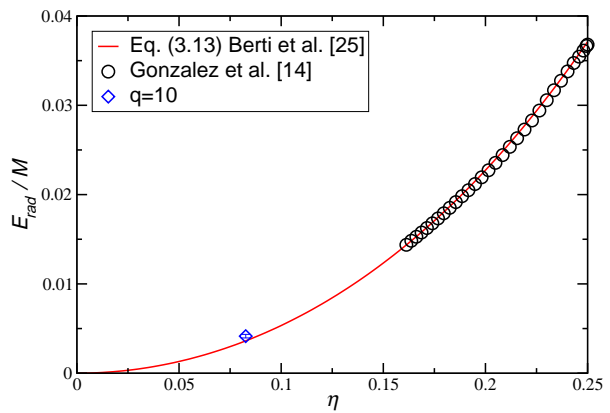}
  \caption{Predictions for the energy radiated during the last three
           orbits (starting at phase 43.3 radians before the
           maximum in the $\ell=2$, $m=2$ mode), merger and ringdown
           are compared
           with numerical data from \cite{Gonzalez2007} as well
           as new results obtained for the mass ratio $q=10:1$.
          }
  \label{fig: E}
\end{figure}
\begin{figure}[b]
  \includegraphics[angle=-90,width=250pt]{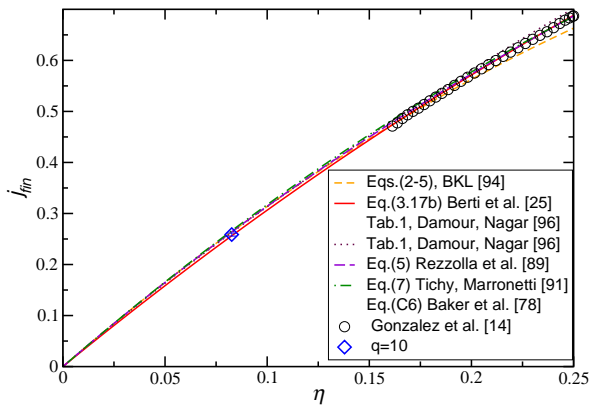}
  \caption{Same as Fig.~\ref{fig: E} but for the final spin.
           The curves resulting from Eq.~(7) of Tichy \& Marronetti
           and from Eq.~(C6) of Baker {\em et al.}
           \cite{Baker2008a, Buonanno2007a} would be indistinguishable
           in this plot and have been represented as the single
           dash-dotted line. The same applies to the curves resulting from
           Berti {\em et al} \cite{Berti2007}
           and the upper predictions of Damour \& Nagar's \cite{Damour2007}
           which assumes the Kepler law (see their Sec.~II).
          }
  \label{fig: jfin}
\end{figure}
\begin{figure}[b]
  \includegraphics[angle=-90,width=250pt]{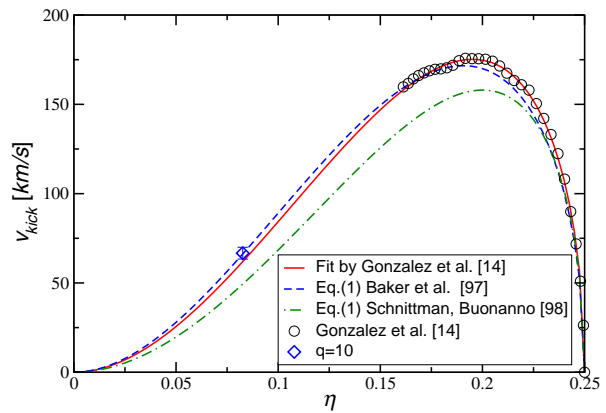}
  \caption{Same as Fig.~\ref{fig: E} but for the recoil.
          }
  \label{fig: recoil}
\end{figure}

In Fig.~\ref{fig: E}, we show the results for the energy
radiated in the form of gravitational waves
during the last three orbits of the inspiral, the merger and the ringdown
as a function of $\eta$. To be precise, we start the integration
at a phase $\Delta \phi=43.3~{\rm rad}$ in the $\ell=2$, $m=2$
mode prior to the maximum amplitude in that mode.
Similar results for mass ratios $1 \le q \le 4$
were presented in \cite{Gonzalez2007}
and approximated with a polynomial fit in Eq.~(3.13) in \cite{Berti2007}.
Our simulations with $q=10$ are in excellent agreement with the
amount of gravitational wave energy expected from the polynomial fit.
We observe similar agreement with the extrapolated prediction
of the peak of the energy flux
given in Baker {\em et al.} \cite{Baker2008a}. Using their
relation (6), we obtain $\dot{E}_{22,{\rm max}} = 3.26\times 10^{-5}$ to
be compared with our numerical result $\dot{E}_{22, {\rm max}}
= 3.18\times 10^{-5}$. The difference is comfortably within either study's
error estimates.

In Fig.~\ref{fig: jfin}, we consider fitting
formulas for the final spin of the merged hole
taken from Berti {\em et al.}~\cite{Berti2007},
Rezzolla {\em et al.}~\cite{Rezzolla2007a}, Tichy \& Marronetti
\cite{Tichy2008} and Buonanno {\em et al.}~\cite{Buonanno2007a}
(Eq.~(C6) in Baker {\em et al.}~\cite{Baker2008a}) as well as the EOB
predictions by Damour \& Nagar \cite{Damour2007}.
The figure demonstrates that all formulas
agree very well with both, the results of our previous study in
\cite{Gonzalez2007}
in the range $\eta=0.16...0.25$ and the new value at $\eta=0.0826$.
On the other hand, we are not able to discriminate between the
different formulas based on our results of non-spinning binaries.
The figure also contains the predicted final spin of the
model proposed by Buonanno, Kidder \& Lehner (BKL) \cite{Buonanno2007}.
The values have been obtained by numerically solving their Eqs.~(2-5).
Given that they do not fit existing numerical data but model
the final spin, their agreement with the numerical results
is remarkable. Even for nearly equal masses, the deviations
are relatively small, of the order of $5~\%$.

Fig.~\ref{fig: recoil} shows a similar analysis for the gravitational
recoil using analytic
predictions by Gonzalez {\em et al.} \cite{Gonzalez2007},
Baker {\em et al.} \cite{Baker2008} and Schnittman \& Buonanno
\cite{Schnittman2007}. We thereby additionally test Eq.~(2)
of Lousto \& Zlochower \cite{Lousto2007} which by construction
reduces to the prediction of \cite{Gonzalez2007} in the
limit of initially non-spinning holes.
We further emphasize that Schnittman \& Buonanno
model the recoil using the effective one body (EOB) method
instead of merely fitting available numerical data (cf.~BKL for the final
spin above). It is natural in this case to expect,
larger deviations from the numerically obtained values.
In analogy to the comparison of the final spin,
we observe generally good agreement between predicted values and the
numerical results including $q=10$.
This agreement is encouraging as we believe it highly
unlikely that there exist local extrema in either curve in the
range $\eta=0.05 \ldots 0.15$ and therefore provides strong support for all
formulas in the limit of vanishing initial spin of the holes.

\subsection{Multipolar structure}
\label{sec: modes}

\begin{figure}[b]
  \includegraphics[angle=-90,width=250pt]{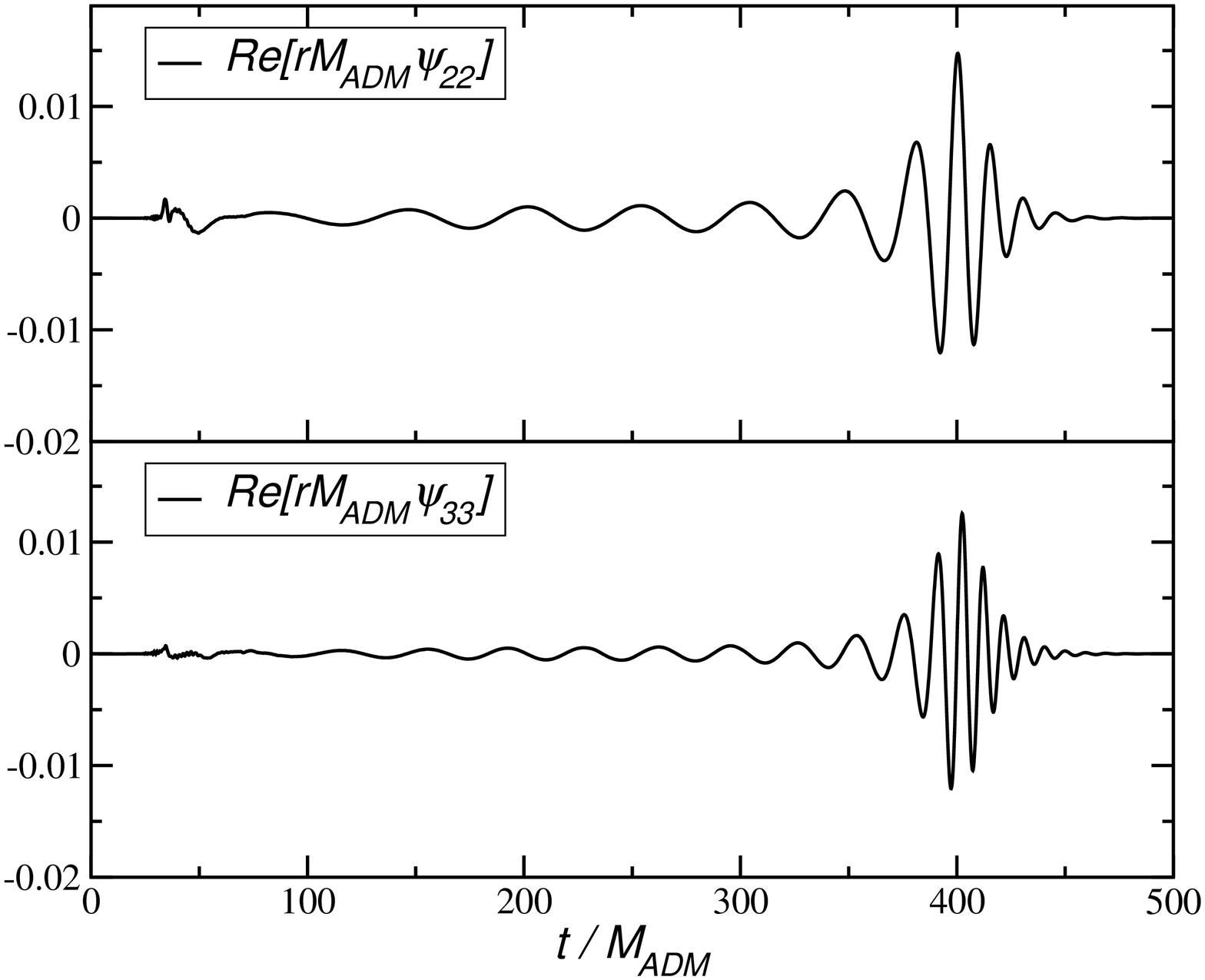}
  \includegraphics[angle=-90,width=250pt]{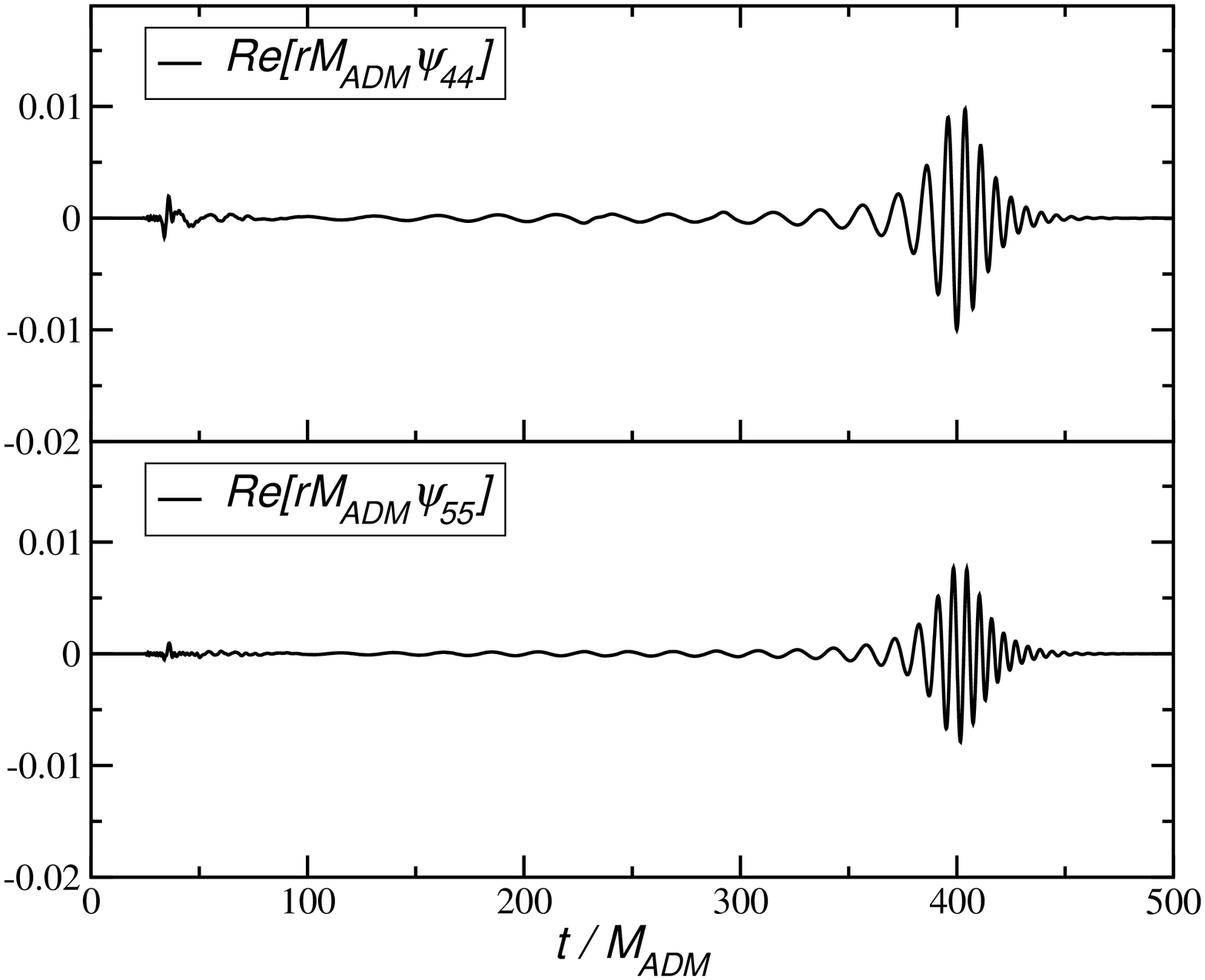}
  \caption{Multipole coefficients $r_{\rm ex}M_{\rm ADM}
           \psi_{\ell m}$ for $\ell=m=2,...,5$
           extracted at $36.5~M_{\rm ADM}$.
          }
  \label{fig: psilm}
\end{figure}

Gravitational waves are commonly decomposed into multipoles using
spherical harmonics of spin weight $-2$. Specifically,
the complex Newman-Penrose
scalar $\Psi_4$ extracted at constant radius $r_{\rm ex}$ is
written as a sum (see e.~g.~\cite{Thorne1980})
\begin{equation}
  \Psi_4(t,\theta,\phi) = \sum_{\ell, m} \psi_{\ell m}(t) {}_{-2}Y_{\ell m}
      (\theta, \phi).
  \label{eq: BSSN_chi}
\end{equation}
For illustration, we show in Fig.~\ref{fig: psilm} the real part of the
multipole coefficients for $\ell=m=2,...,5$.
Most of the earlier studies of black-hole binaries focused on non-spinning,
equal-mass systems where $>98\%$ is radiated in the quadrupole terms
$\ell=2$. For more general classes of binaries, however, a significant
fraction of the gravitational wave energy is radiated in higher order
modes, e.g.\ \cite{Berti2007}. Simultaneously, the gravitational wave signal
predicted for a given orientation of the binary's orbital plane
will contain substantial contributions from higher order multipoles
and may thus exhibit a pattern much more complex than visible in the
quadrupole amplitude $\psi_{2\pm 2}(t)$. While this complex multipolar
structure places higher demands on the modeling of the gravitational
wave sources, it provides us with a large amount of information
in the effort to estimate parameters from gravitational wave observations.

A comprehensive study of exploiting such information of higher order
multipoles in the context of gravitational wave
data analysis is beyond the scope of this paper. In this section we
therefore restrict ourselves to a discussion of the multipolar distribution
of the gravitational wave energy
and merely illustrate the significance of including higher order modes
in the waveform for fixed inclination angle of the orbital plane relative
to an observer.

The amount of energy contained in different multipoles for the inspiral
of binaries with mass ratio $q=1...4$ has been given in Table IV of
\cite{Berti2007}.

\begin{figure}[b]
  \includegraphics[angle=-90,width=250pt]{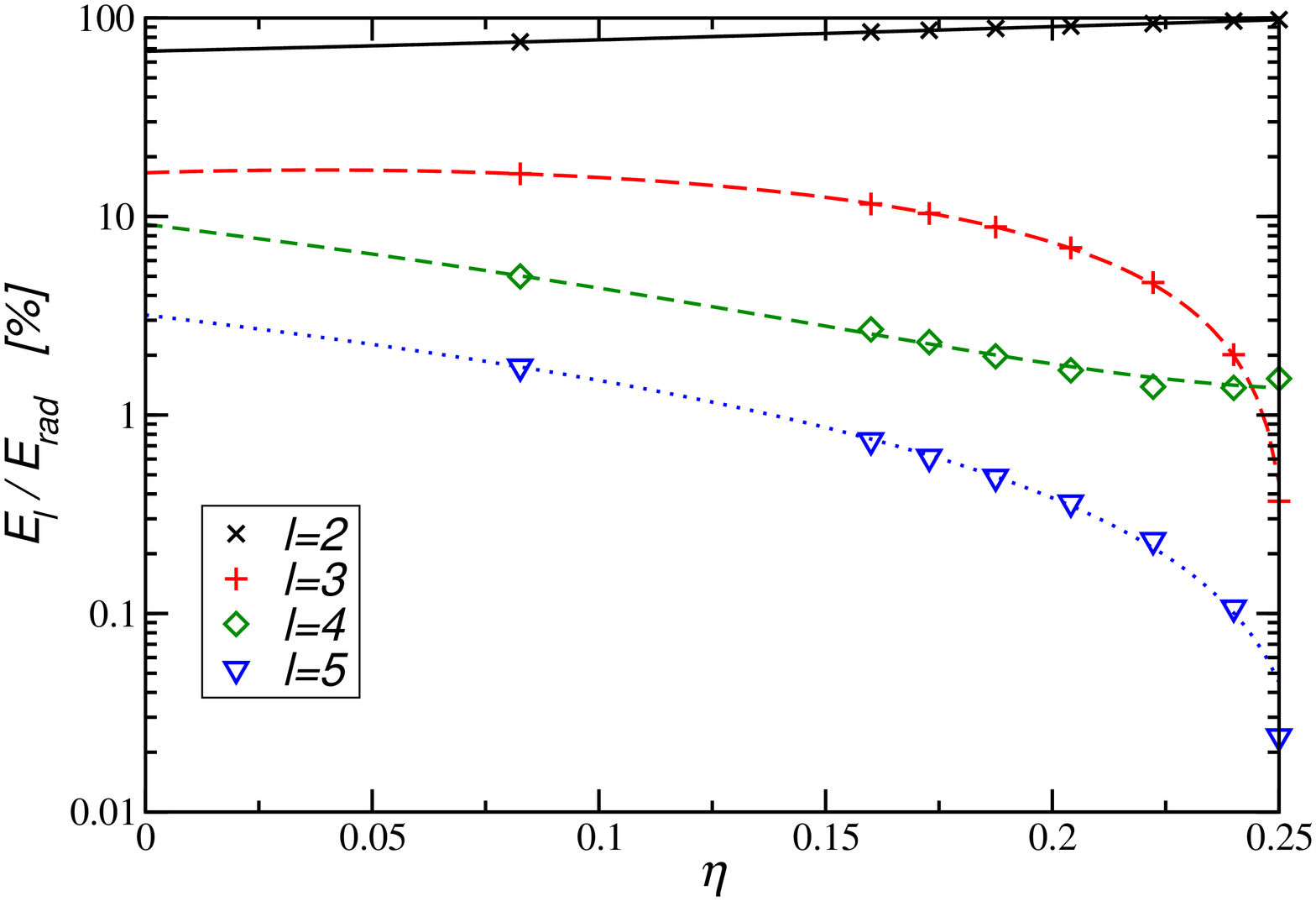}
  \includegraphics[angle=-90,width=250pt]{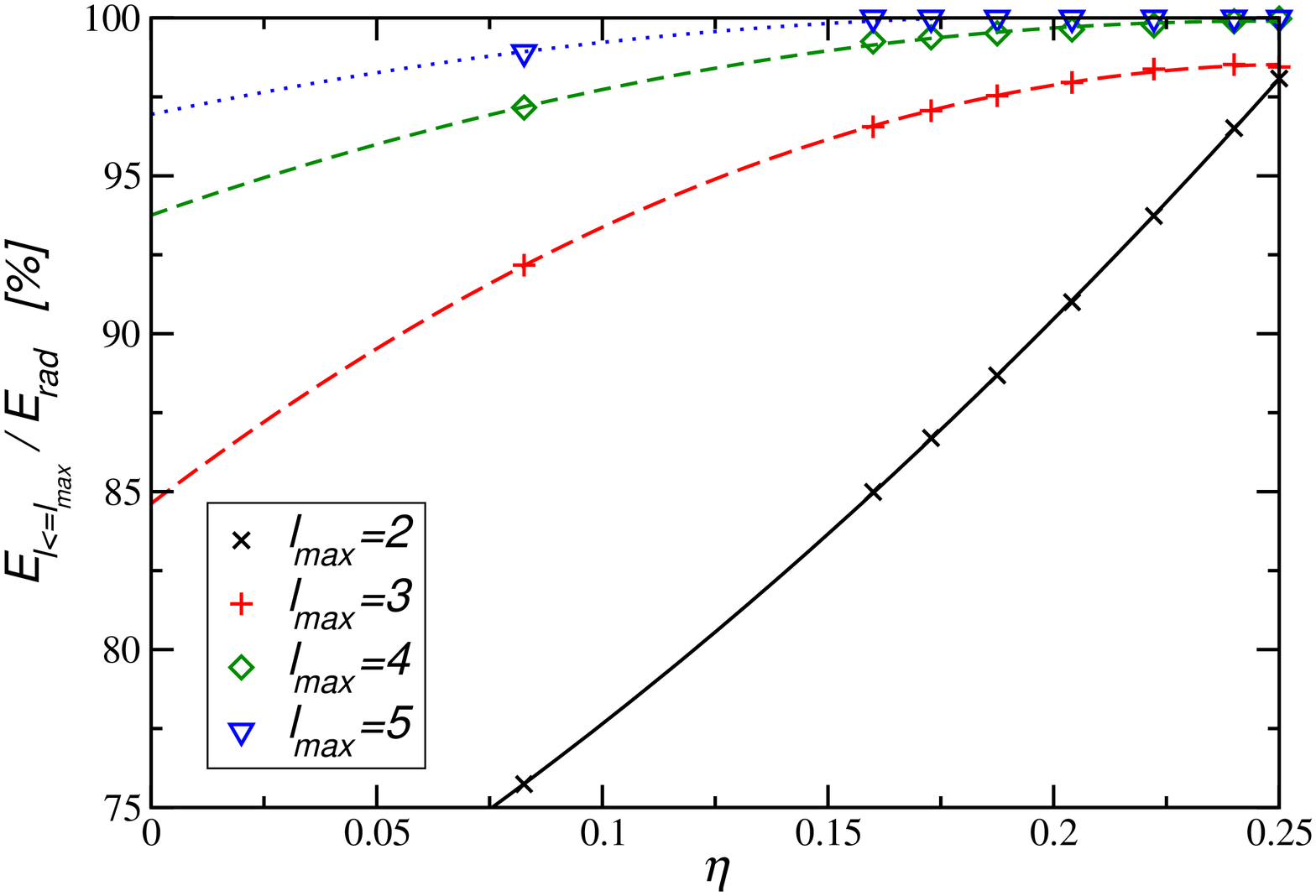}
  \caption{The energy contained in the individual multipoles
           $\ell=\rm const$ (upper panel) as well as the
           energy radiated in all modes up to a given
           $\ell_{\rm max}$ (lower panel) as a function of the dimensionless
           mass parameter $\eta$.
          }
  \label{fig: Emodes}
\end{figure}

We graphically display the energy contained in the multipoles in
Fig.~\ref{fig: Emodes}. Here, the upper panel displays the energy contained
in the multipoles corresponding to a particular value of $\ell$,
whereas the symbols in the lower panel show the fractional energy
contained in all modes up to $\ell \le \ell_{\rm max}$. We discard
the first $50~M_{\rm ADM}$ of the waveform which is dominated by
spurious radiation inherent to the initial data. This corresponds
to the radiated energy starting at phase $43.3~{\rm rad}$ prior to the
maximum in the $\ell=2$, $m=2$ mode.
The resulting energy is well approximated by quadratic polynomials.
Specifically, we obtain the following fits for the numerical
data in the range $1\le q\le 10$ corresponding to $0.25 \ge \eta\ge0.0826$
\begin{eqnarray}
  E_{\ell =2}/E_{\rm rad} &=& 68.0 + 80.4\eta + 159.2 \eta^2, \\[10pt]
  E_{\ell \le 3}/E_{\rm rad} &=& 84.6 + 108.8 \eta - 212.7 \eta^2, \\[10pt]
  E_{\ell \le 4}/E_{\rm rad} &=& 93.8 + 49.9\eta - 101.3\eta^2, \\[10pt]
  E_{\ell \le 5}/E_{\rm rad} &=& 96.9 + 30.0\eta - 72.1\eta^2.
\end{eqnarray}
The trend for higher order multipoles to carry a larger fraction
of the total radiated energy $E_{\rm rad}$
is clearly maintained for $q=10$. Close
inspection of $E_{\ell =3}$ reveals a local maximum near $\eta=0.04$.
We believe this to be an artifact of the limited accuracy of the
data and the polynomial fits.

The significance of higher multipoles also reveals itself in the
gravitational wave signal observed at fixed values for the inclination
angle $\theta$ of the orbital plane of the binary. The combination of
all multipoles in the signal $\Psi_4$ shows
a much more complex structure than the quadrupole on its own.
\begin{figure}[b]
  \includegraphics[angle=-90,width=250pt]{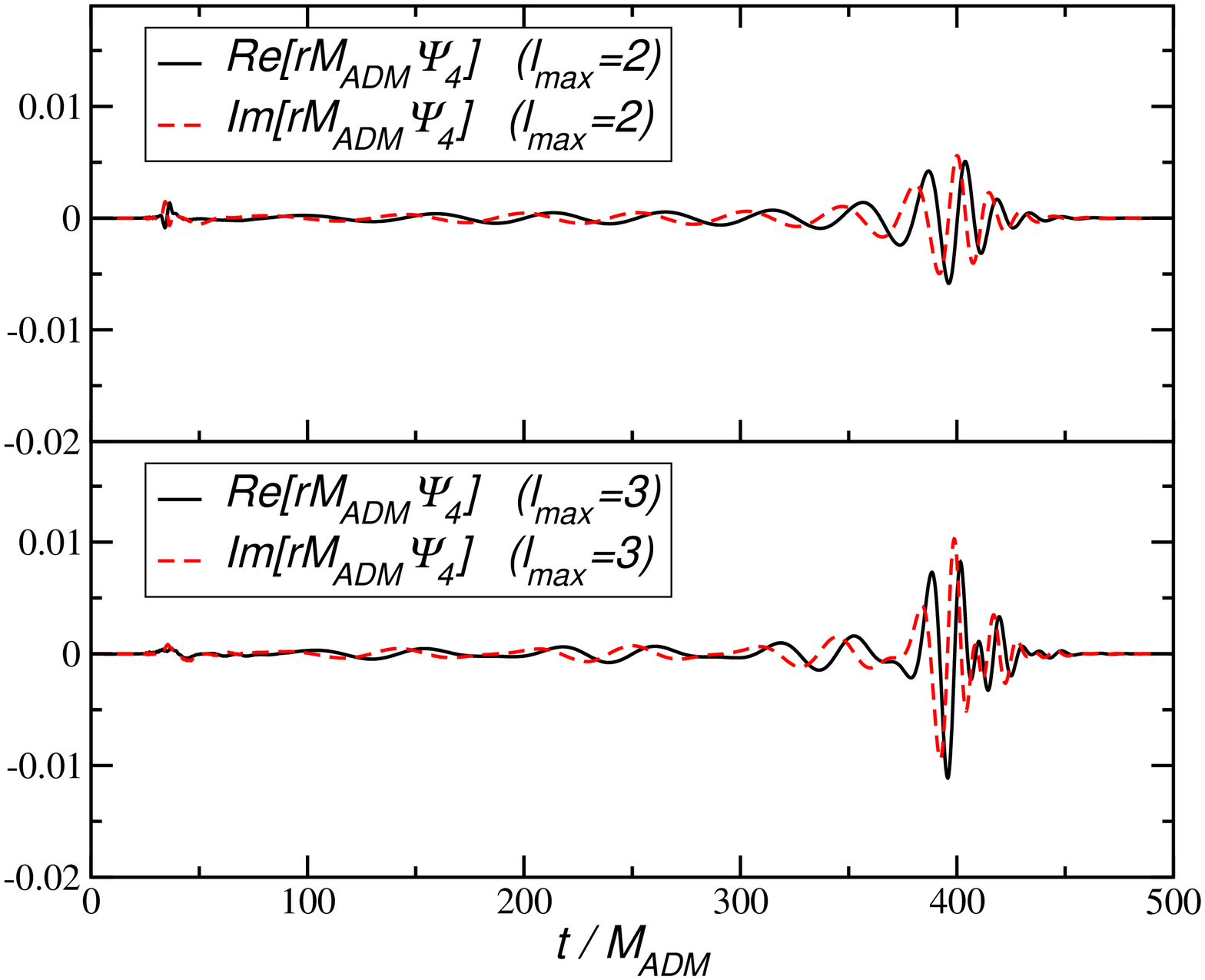}
  \includegraphics[angle=-90,width=250pt]{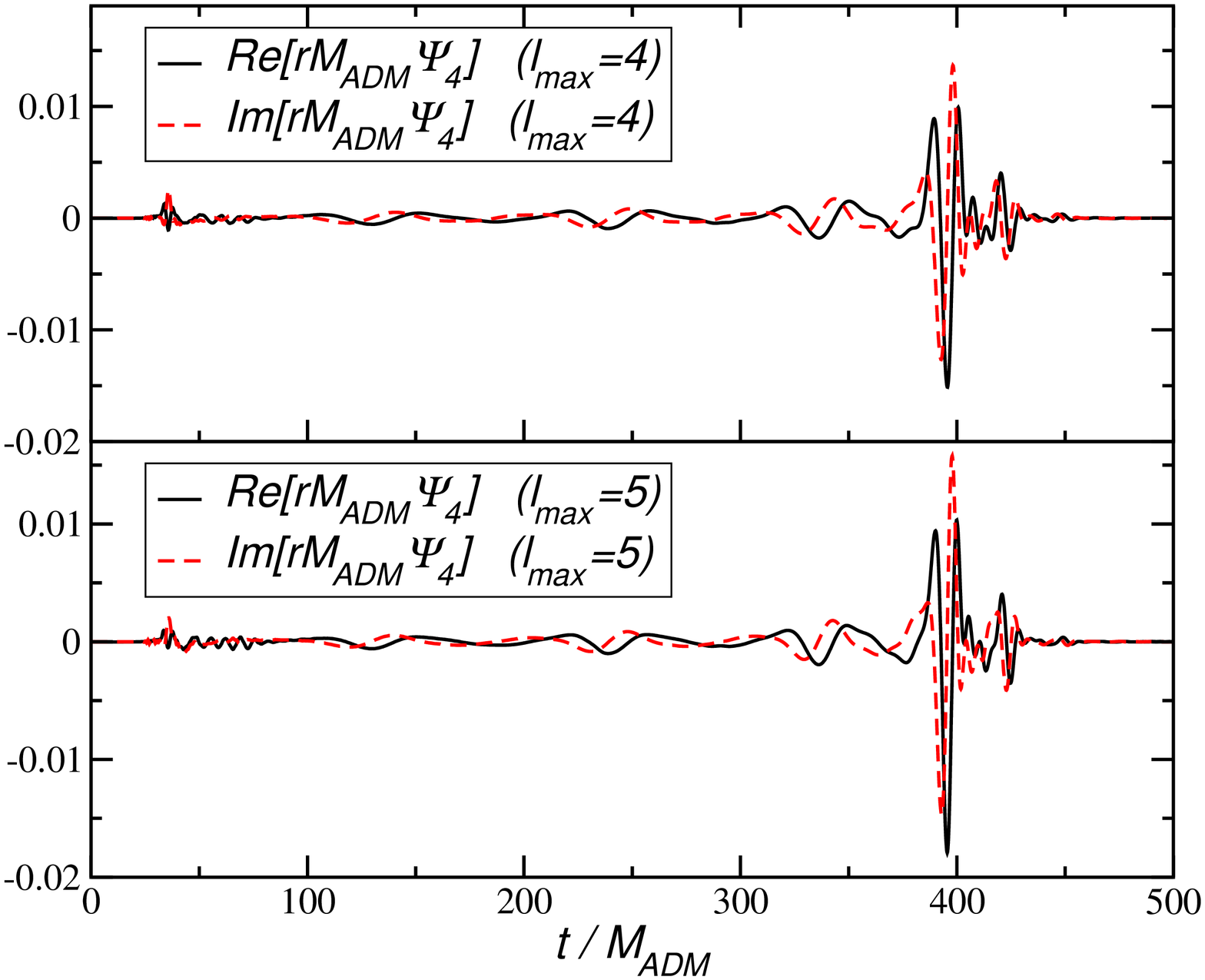}
  \caption{The Newman-Penrose scalar $\Psi_4$ obtained at
           an inclination of the
           orbital plane of $\theta=45.5$ degrees including
           multipoles up to and including $\ell_{\rm max}=2$
           , $3$, $4$ and $5$ (from top to bottom).
          }
  \label{fig: psi4}
\end{figure}
We illustrate this in Fig.~\ref{fig: psi4} where we plot real and
imaginary part of $\Psi_4$ at $r_{\rm ex}=36.5~M_{\rm ADM}$ and
$\theta=45.5^{\circ}$. The inclination angle has been chosen
somewhat arbitrarily, but we get the same general result for
other choices of $\theta$.
The pattern resulting from the inclusion of the octupole and higher
order modes
in the bottom panels of Fig.~\ref{fig: psi4} shows a non-monotonic
increase in the local maxima of the wave signal during inspiral
which is remniscent of the pure quadrupole signal in eccentric
inspirals (cf.~the ``MayaKranc e02'' entry in Fig.~1 of
Ref.~\cite{Aylott2009}).
In contrast there is little evidence of eccentricity
if we consider the individual modes of our simulation
in Fig.~\ref{fig: psilm}. As a further impact of the higher
order multipoles we note
the significant increase in the amplitude between the top and bottom
panel of Fig.~\ref{fig: psi4}.

These observations qualitatively illustrate the
importance of including higher order multipoles in gravitational
wave data analysis as has been pointed out quantitatively
in the literature
\cite{VanDenBroeck2006, VanDenBroeck2006a, Arun2007, Arun2007a, Berti2007,
Pan2007, Porter2008, Baker2008a}\footnote{Higher order harmonics as
discussed in some of these references are not strictly equivalent
to higher order multipoles but often imply the inclusion of additional
multipoles}.
In future work, we plan to investigate this issue more systematically
using our numerical data in the context of matched filtering.

\section{Conclusions}
\label{sec: conclusions}

In this work, we have pushed the mass ratio in numerical simulations
of inspiralling black-hole binaries significantly
beyond what has previously been published. Our simulations have been
demonstrated to be convergent consistently with the discretization
properties of the numerical code. We have resolved an issue specific to
larger mass ratios by choosing a specific value of the shift damping
parameter.
The overall errors in wave amplitude
and phase as well as radiated energy and momenta are about $5\%$. We
have thus been able to validate existing fitting formulas for the
amount of energy and linear momentum as well as the final spin of
the merging hole in a range of the mass ratio previously unexplored.
Within the error bars, we find our numerical results in agreement
with phenomenological formulas. These are of high importance in
the modeling of astrophysical phenomena, such as
the growth of supermassive black holes. At least
in the case of non-spinning holes, we conjecture that the fitting
formulas can be used over the entire range of the dimensionless
mass ratio parameter $\eta$.

Extending previous work \cite{Berti2007}, we have further demonstrated
that the percentage of gravitational wave energy radiated in higher
order multipoles increases significantly as the mass ratio
deviates from the equal-mass limit $\eta=0.25$ or $q=1$. Quadratic
fitting of our numerical data indicates that about $32\%$ of the total
radiated energy will be contained in $\ell>2$ as the
extreme mass ratio (EMR) limit is approached. For the $q=10$ case
at hand, we find about $25\%$ of the energy to be contained in modes
higher than the quadrupole. This distribution of the energy
also manifests itself in the shape of the gravitational waveform
as measured for a fixed inclination of the binary orbit relative to
the observer. The case $\theta\approx 45$ degrees exhibits a significant
change in the wave signal as we include higher order multipoles. While
we only display results for one value of $\theta$, we find this
behavior to be similar for arbitrary alternative inclinations.

It will be important to extend the current study to spinning binaries
in the future. This will allow us to address various important questions
in astrophysics and gravitational wave physics. For example, there
remain uncertainties about the magnitude of the recoil effect
for spinning binaries of unequal mass
\cite{Campanelli2007, Baker2008, Lousto2008}.
Also, it will be
important to compare fitting formulas for final spin and recoil
with numerical results for initially spinning black holes.

The use
of numerical waveforms in gravitational wave data analysis further
requires careful comparisons with post-Newtonian predictions and
the combination of numerical with post-Newtonian waveforms
(e.~g.~\cite{Baker2006c, Pan2007, Hannam2007, Boyle2007, Gopakumar2007, 
Damour2008, Hinder2008, Campanelli2008, Boyle2008}).
In future work, we plan to perform such comparisons including various
post-Newtonian techniques. We also believe that the current
simulations facilitate approximate comparisons with perturbative calculations
of EMR binaries.
Finally, an intriguing question concerns the
properties of black hole collisions or scattering experiments
involving unequal-mass binaries traveling close to the speed
of light. Studies have so far been restricted to collisions
of nonspinning equal-mass binaries \cite{Sperhake2008, Shibata2008}
and predict
that as much as $\sim 14\%$ of the total energy of the system can
be radiated in head-on collisions and even larger quantities for
non-zero impact parameter.

In summary, accurate simulations of black-hole binaries with mass
ratio $q=10$ are not only possible using current numerical techniques
(if somewhat expensive), but also reveal a richness in structure
beyond what has been observed in the nearly equal mass case. We consider our
simulations of the non-spinning case to be the starting point of more
exhaustive studies involving spins and/or eccentricity, which will be
of significant
value for current efforts to observe gravitational waves and improve
our understanding of astrophysical questions involving the merger of
black-hole binaries.

\begin{acknowledgments}
It is a pleasure to thank Emanuele Berti, Vitor Cardoso,
Mark Hannam, Sascha Husa, and Doreen M\"uller for discussions.
This work was supported in part by DFG grant SFB/Transregio~7
``Gravitational Wave Astronomy'', the DLR (Deutsches Zentrum f\"ur Luft
und Raumfahrt), by grants from the Sherman Fairchild
Foundation to Caltech, by NSF grants PHY-0601459, PHY-0652995
and PHY-090003, by
grant CIC 4.23 to Universidad Michoacana, PROMEP UMICH-PTC-210 and
UMICH-CA-22 from SEP M\'exico and CONACyT grant number 79601.
Computations were performed on the HLRB2 at LRZ Munich. We acknowledge
support from the ILIAS Sixth Framework programme.
\end{acknowledgments}

%\bibliography{uli.bib}

\end{document}